\documentclass[%
 aip,
 amsmath,amssymb,
 reprint,%
]{revtex4-1}

\usepackage{graphicx}% Include figure files
\usepackage{dcolumn}% Align table columns on decimal point
\usepackage{hyperref}
\hypersetup{
    colorlinks=true,
    linkcolor=black,
    filecolor=black,      
    urlcolor=black,
    pdftitle={Overleaf Example},
    pdfpagemode=FullScreen,
    }
\usepackage{bm}% bold math
\usepackage[utf8]{inputenc}
\usepackage[T1]{fontenc}
\usepackage{mathptmx}
\usepackage{cleveref}
\usepackage{enumitem}
\usepackage{bbm}
\usepackage{etoolbox}
\usepackage{titlesec}
\usepackage{color,soul}

\titlespacing\section{0pt}{12pt plus 4pt minus 2pt}{5pt plus 2pt minus 2pt}
\titlespacing\subsection{0pt}{12pt plus 4pt minus 2pt}{3pt plus 2pt minus 2pt}

\newcommand{\bfx}{\mathbf{x}}
\newcommand{\bfz}{\mathbf{z}}

\newcommand{\bftheta}{\boldsymbol\theta}
\newcommand{\revision}{\textcolor{black}}

\DeclareMathOperator*{\argmin}{arg\,min}

\makeatletter
\def\@email#1#2{%
 \endgroup
 \patchcmd{\titleblock@produce}
  {\frontmatter@RRAPformat}
  {\frontmatter@RRAPformat{\produce@RRAP{*#1\href{mailto:#2}{#2}}}\frontmatter@RRAPformat}
  {}{}
}%
\makeatother
\begin{document}

\preprint{AIP/123-QED}

\title{Knowledge-Based Learning of Nonlinear Dynamics and Chaos}
% Force line breaks with \\
\author{Tom Z. Jiahao}
\email{zjh@seas.upenn.edu}
\affiliation{Department of Computer and Information Science, \\University of Pennsylvania, Philadelphia, PA, 19104, USA}

\author{M. Ani Hsieh}
\affiliation{Department of Mechanical Engineering and Applied Mechanics, \\University of Pennsylvania, Philadelphia, PA, 19104, USA}

\author{Eric Forgoston}
\affiliation{Department of Applied Mathematics and Statistics,\\ Montclair State University, Montclair NJ, 07043, USA}

\date{\today}% It is always \today, today,

\begin{abstract}
Extracting predictive models from nonlinear systems is a central task in scientific machine learning. One key problem is the reconciliation between modern data-driven approaches and first principles. Despite rapid advances in machine learning techniques, embedding domain knowledge into data-driven models remains a challenge. In this work, we present a universal learning framework for extracting predictive models from nonlinear systems based on observations. Our framework can readily incorporate first principle knowledge because it naturally models nonlinear systems as continuous-time systems. This both improves the extracted models' extrapolation power and reduces the amount of data needed for training. In addition, our framework has the advantages of robustness to observational noise and applicability to irregularly sampled data. We demonstrate the effectiveness of our scheme by learning predictive models for a wide variety of systems including a stiff Van der Pol oscillator, the Lorenz system, and the Kuramoto-Sivashinsky equation. For the Lorenz system, different types of domain knowledge are incorporated to demonstrate the strength of knowledge embedding in data-driven system identification.
\end{abstract}

\maketitle

\begin{quotation}
We consider the general problem of data-driven modeling of dynamical systems from past observations. We seek to address two important questions: (i) how to use machine learning to model general classes of dynamical systems, especially those with nonlinear and chaotic dynamics, and (ii) how to reconcile data-driven models and first principles knowledge. We propose Knowledge-Based Neural Ordinary Differential Equations (K-NODE), which, unlike other machine learning techniques, is marked by its flexibility for first principles knowledge incorporation, and applicability to a wide variety of dynamical systems. Furthermore, our framework is robust to observational noise and irregularly sampled data. Given the abundance of simulated and real data available to scientists and engineers, our framework can be used to extract meaningful correlations and identify system relationships. The proposed framework enables acquisition of new physical insights and improves understanding of a wide variety of complex nonlinear systems in diverse scientific disciplines.
\end{quotation}

\section{Introduction} 

Recent advances in machine learning and data analytics have largely been fueled by the vast amount of data and have resulted in significant advances in various scientific disciplines~\cite{MLTrend, DataTrend}. Deep learning tools such as recurrent and convolutional neural networks have enabled the identification of coherent data patterns commonly imperceptible to humans. However, most of these data-driven techniques perform poorly when they are trained with insufficient data or used to extrapolate beyond the sampled data. These deficiencies are largely due to the inability to incorporate first principles domain knowledge. By combining first principles with deep learning models, knowledge-based learning can potentially address these challenges by leveraging the extrapolation power of first principles knowledge. However, most existing deep learning models are incompatible with these knowledge.

Since the emergence of calculus, differential equations have been successfully used to model real world phenomena from planetary motions, fluid processes, to biological systems. 
Differential equations are compact representations of vector fields on which the evolution of dynamical systems can be realized. Fundamental to the idea of differential equations is the assumption that time and space are continuous.  Our vast pool of interpretable first principles knowledge are built upon these underlying assumptions. In contrast, many data-driven approaches for modeling dynamics do not assume continuity for systems and are fundamentally discrete in nature.

One of the earliest data-driven approaches to model dynamical systems builds on Takens' embedding theorem and represents systems as time delay models by embedding the original system's states into delayed snapshots~\cite{Takens80, PaduartPoly, Wanfeedforward}. In comparison, modern machine learning strategies employ recurrent neural networks (RNNs) which have memory properties as a result of feedback loops. Both long-short term memory (LSTM)~\cite{LSTM} and reservoir computers (RCs)~\cite{EchoState} are examples of RNNs that have been employed to predict two-dimensional fluid flows~\cite{qraitem2020bridging} and to describe the chaotic dynamics of a one-dimensional Kuramoto-Sivashinsky model~\cite{OttRescomp}. However, both time delay models and RNN-based approaches only output the solutions of dynamical systems at prescribed time intervals. Although the solutions may appear continuous, the continuity of the original system is inevitably lost. In fact, this lack of model continuity is one of the reasons for the incompatibility between knowledge and data-driven models. There have been attempts to combine knowledge with reservoir computers~\cite{OttRescomp} for modeling spatiotemporally chaotic systems. Although this hybrid approach showed improved performance, only the solutions of a first principle model were incorporated instead of the model itself. Consequently, the reservoir computer and knowledge are still disjoint.

Most recently, there has been an increased focus in developing strategies that explicitly extract differential equations of a system to represent its dynamics. One method uses sparse regression to determine the combination of basis functions that best describes the vector field from observations~\cite{SINDy, EntropicRegression}, and therefore is able to extract interpretable mathematical models. Nevertheless, a fundamental challenge with this method is the need for a predetermined library of basis functions. If the correct terms are missing from the library, the resulting models may be inadequate at describing the system dynamics. In addition, sparse regression does not scale with high-dimensional systems as the library of functions would become very large.

Chen {\it et al.}  introduced a new family of neural networks, commonly referred to as neural ordinary differential equations (NODE)~\cite{conf/nips/ChenRBD18}. NODE and its variants~\cite{pmlr-v108-li20i, Zang_2019, ayed:learning_partial, NIPS2019_8577} combine the power of modern machine learning with the formalism of differential equations, and are fundamentally continuous \cite{conf/nips/ChenRBD18}. \revision{Separately, a scientific machine learning library was built using NODE to incorporate physical constraints into data-driven models.  This framework has demonstrated how to incorporate a wide range of knowledge into scientific machine learning\cite{rackauckas2020universal}. Nonetheless, existing NODE formulations assume stable dynamics for the systems they model in order to achieve convergence~\cite{curseofsensitivity}. As a result, these and other NODE variants have yet to generalize to systems with unstable or temporally chaotic dynamics.}

In this paper we introduce K-NODE, a novel universal framework for modeling system dynamics from observations. Our proposed framework is universal in that it generalizes the learning task to any continuous-time dynamical system including those with unstable and chaotic behaviors. It is also robust to noisy and irregularly sampled data, and can scale to high-dimensional systems. Most importantly, our framework can readily incorporate first principles domain knowledge into neural networks, significantly improving their extrapolation ability. We redefine the system identification task as a constrained optimization problem which converges regardless of the stability of systems. Similar to NODE, the optimization problem can be efficiently solved using the adjoint sensitivity method~\cite{Cao2003AdjointSA}, which allows for constant-memory gradient propagation regardless of the number of interpolation steps between the observation time intervals.

We will first demonstrate the effectiveness of our framework by modeling a variety of nonlinear systems using only neural networks. Then we will show how different forms of knowledge can be incorporated to significantly reduce the amount of data needed for training, and to improve the extrapolation power of the models beyond the sampled data.

\section{Knowledge-based Neural ODEs (K-NODE)}
\begin{figure}
    \centering
    \includegraphics[width=0.38\textwidth]{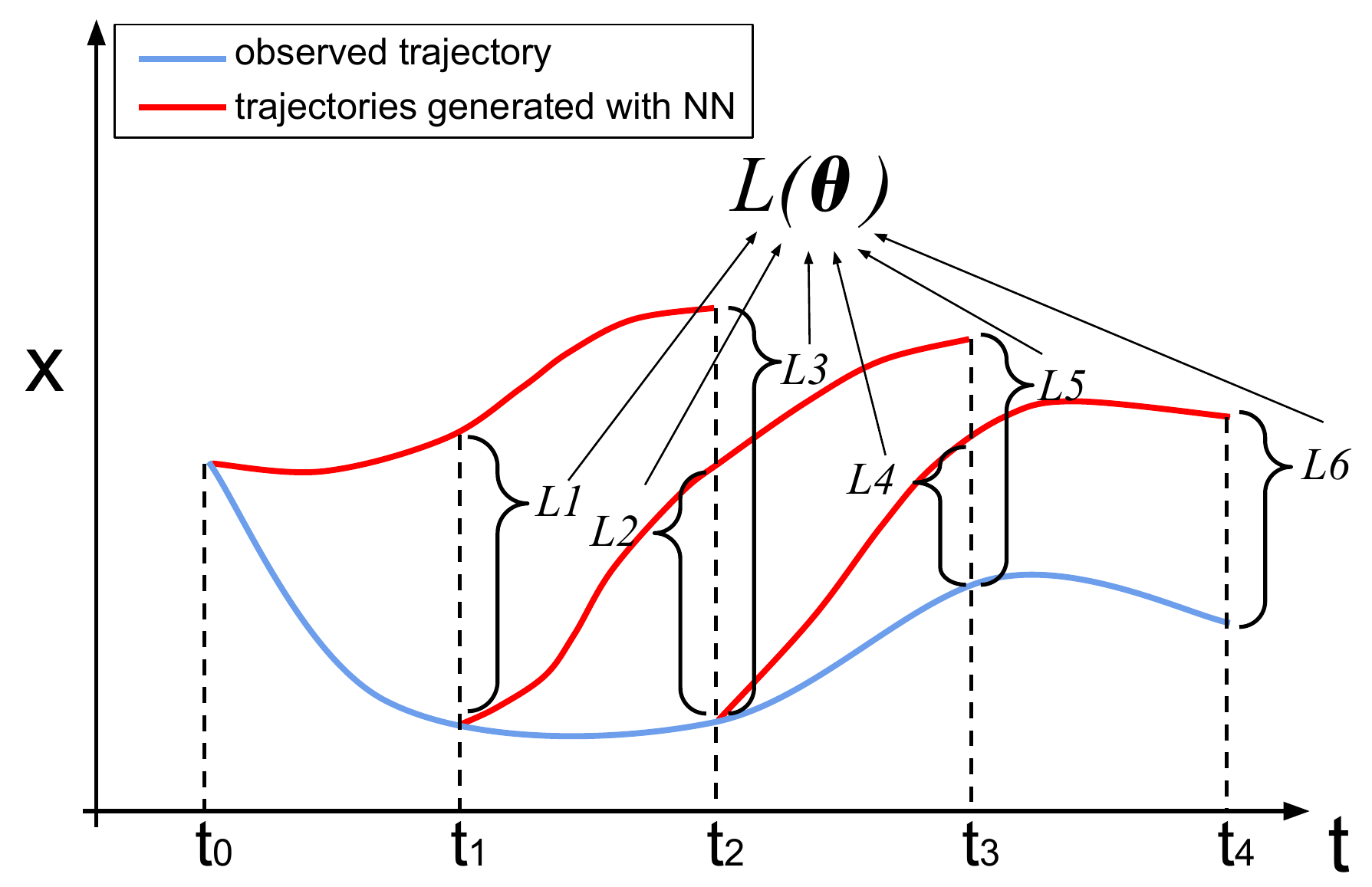}
    \caption{A 1D example of the constrained optimization problem. The blue curve is the observation, and the red curves are the trajectories generated by the neural network $\hat{f}$. In this example $\alpha = 2$, and therefore the neural network generates a trajectory of length 2 beginning with the sampled state at every time step. The loss is then computed between the blue and red curves at every sampled time steps.}
    \label{fig: constrained_opt}
\end{figure}

{We are given $m$ observations of the trajectory generated by a dynamical system sampled at $T = \{t_1, t_2,...,t_m\}, t_i\in \mathbb{R}$:}
\begin{equation*}
\mathbf{Z}=\left[\begin{array}{c}
(\bfz)^\top\left({t_1}\right) \\
(\bfz)^\top\left({t_2}\right) \\
\vdots \\
(\bfz)^\top\left({t_m}\right)
\end{array}\right]=\left[\begin{array}{cccc}
z_{1}\left({t_1}\right) & z_{2}\left({t_1}\right) & \cdots & z_{n}\left({t_1}\right) \\
z_{1}\left({t_2}\right) & z_{2}\left({t_2}\right) & \cdots & z_{n}\left({t_2}\right) \\
\vdots & \vdots & \ddots & \vdots \\
z_{1}\left({t_m}\right) & z_{2}\left({t_m}\right) & \cdots & z_{n}\left({t_m}\right)
\end{array}\right],
\end{equation*}
where $\mathbf{Z}\in \mathbb{R}^{m\times n}$ is the matrix containing the observations, and the vector $\bfz(t_i)\in \mathbb{R}^{n}$ is the observation of the state at $t_i$. {Assume the true model of this dynamical system is given by
\begin{equation*}
    \dot{\bfx} = f\left(\bfx, t\right),
\end{equation*}
where $\bfx(t)\in \mathbb{R}^n$ is the $n$-dimensional state vector at time $t$. Let $\tilde{f}$ denote the model based on our current understanding of the system. Then $\tilde{f}$ is given by
\begin{equation}
    \dot{\bfx} = \tilde{f}(\bfx, t).
    \label{eqn: imperfect model}
\end{equation}}

{We say $\tilde{f}$ is the \textit{knowledge} we have about the system. This knowledge may be perfect, partially known, and/or partially correct, \textit{i.e.}, imperfect -- either lacking the correct nonlinear terms or not having the correct steady state behavior of the true model. Given this knowledge $\tilde{f}$, our goal is to approximate the function $f$ with a hybrid model $\hat{f}(\bfx, t, \tilde{f}(\bfx, t), \bftheta)$, which incorporates the knowledge $\tilde{f}$, and is parameterized with the vector $\bftheta$. In this work, we represent $\hat{f}$ using a combination of artificial neural networks and the knowledge $\tilde{f}$.}

\begin{figure}
    \centering
    \includegraphics[width=0.4\textwidth]{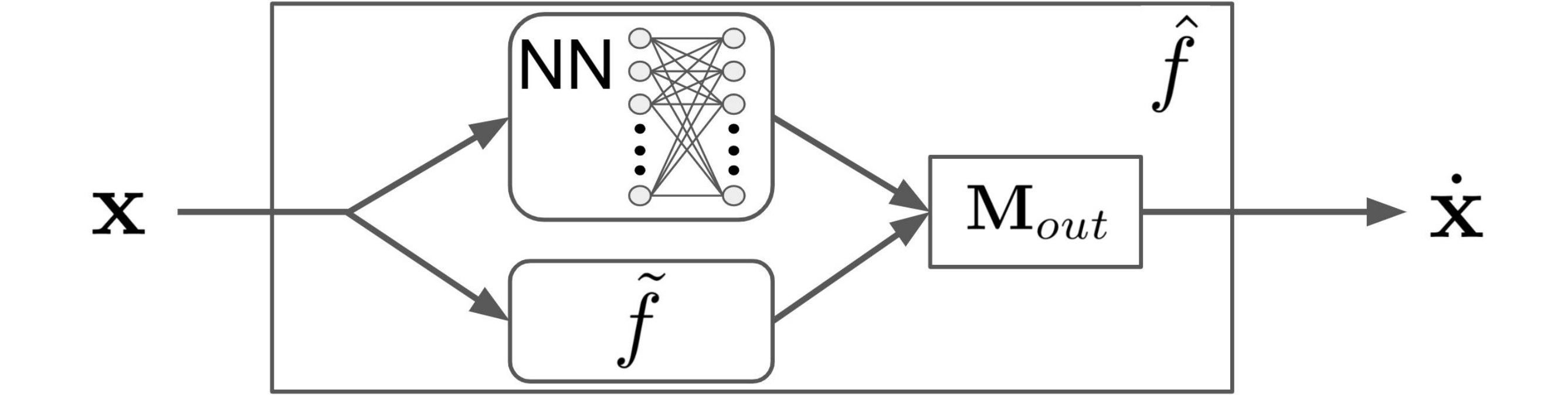}
    \caption{A hybrid architecture incorporating an imperfect vector field $\tilde{f}$ into the model $\hat{f}$. Outputs from the neural network and the knowledge block $\tilde{f}$ are then linearly coupled.}
    \label{fig: hybrid_architecture}
\end{figure}

{While it is flexible how $\tilde{f}$ gets incorporated into $\hat{f}$, in this work we consider linearly coupling the outputs from $\tilde{f}$ and the neural network using a matrix $\mathbf{M}_{out}\in\mathbb{R}^{p\times n}$ with output biases. Both $\mathbf{M}_{out}$ and the biases are co-trained with the neural network. Given the input size $p$, $\mathbf{M}_{out}$ and the biases are initialized uniformly random from $[-1/\sqrt{p}, 1/\sqrt{p}]$. The hybrid architecture of $\hat{f}$ is illustrated in Fig. \ref{fig: hybrid_architecture}.} \revision{Note that our approach to knowledge incorporation differs fundamentally from those of hybrid reservoir computers \cite{OttRescomp}. While hybrid reservoir computers first perform numerical integration for both the knowledge and its reservoirs separately, and then linearly couple their respective solutions, K-NODE first linearly couples the vector fields and then performs numerical integration. K-NODE is able to directly couple the vector field owing to its explicit nature in modeling differential equations.}

We then pose this system identification task as the following constrained optimization problem:
{
\begin{equation}
\begin{aligned}
\min_{\bftheta} \quad & L(\bftheta) \\
\textrm{s.t.} \quad & \dot{\bfx} = \hat{f}(\bfx, t, \tilde{f}(\bfx, t), \bftheta), \\
\quad & \bfx(t_s,\bfz(t_s)) = \bfz(t_s), t_s\in T, 
\end{aligned}
\label{eqn: optimization}
\end{equation}
}
where the first constraint is the differential equation defined by {the hybrid model $\hat{f}$}, the second constraint specifies the initial conditions, and the parameters $\bftheta$ can then be estimated by
    $\bftheta = \argmin_{\bftheta} L(\bftheta).$

For the objective function in \eqref{eqn: optimization}, we define an $L^2$ loss function between the observed trajectory and the trajectory generated by $\hat{f}$ given by
\begin{multline}
    L(\bftheta) = \frac{1}{m-\alpha}\sum^{m-\alpha}_{i=1} \frac{1}{\alpha}\int^{t_{i+\alpha}}_{t_i}\delta(t_s-\tau)\|\bfx(\tau, \bfz(t_i))\\
    - \bfz(\tau)\|^2d\tau,
    \label{eqn: objective}
\end{multline}
where $t_s\in T$ is any sampling time point, and $\delta$ is the Dirac delta function. In general, for any given initial condition, a trajectory can be generated from $\hat{f}$ using a suitable numerical integration scheme. Thus, $\bfx(\tau, \bfz(t_i))$ in $L(\bftheta)$ is defined as the state at time $\tau$ generated by $\hat{f}$ with the initial condition $\bfx(t_i,\bfz(t_i)) = \bfz(t_i)$ at time $t_i$ given by
{
\begin{equation}
    \bfx(\tau, \bfz(t_i)) =  \bfz(t_i) + \int^{\tau}_{t_i}\hat{f}\left(\bfx\left(\omega, \bfz(t_i)\right), \omega, \tilde{f}(\bfx, \omega), \bftheta\right) d\omega.
    \label{eqn: diffeomorphism}
\end{equation}
}
Note that the state at $t_{j+1}$ generated by $\hat{f}$ with the initial condition $\bfz(t_j)$ is called a one-step-ahead diffeomorphic flow associated with $\hat{f}$. In our loss function given by \eqref{eqn: objective}, the integral from $t_i$ to $t_{i+\alpha}$ requires an $\alpha$-step-ahead diffeomorphism associated with $\hat{f}$, and we call $\alpha$ the \textit{lookahead}. While some existing work assumes $\alpha=1$~\cite{SINDy, raissimultistep, ouala2019learning}, we have found that using a larger lookahead can sometimes yield better results. Hence, we treat $\alpha$ as a hyperparameter to be tuned. This constrained optimization formulation is illustrated with a 1D example in Fig. \ref{fig: constrained_opt}. Formulation similar to \eqref{eqn: optimization} has been used to learn chaotic systems using models which heavily incorporates physics but the optimization task is not explicitly stated \cite{NPDE}. \revision{In the original NODE formulation, the loss is defined over the entire trajectory predicted using a neural network. We note that this formulation is one of the main contributors for the numerical instability when learning chaotic systems, since chaotic trajectories are deemed to diverge exponentially fast in time. While there is no theoretical guarantee for convergence using our optimization formulation in \eqref{eqn: optimization}, our empirical results have shown that our formulation is capable of achieving convergence for systems with chaotic or unstable dynamics.}

While the optimization problem in \eqref{eqn: optimization} can be solved using the conventional backpropagation, the adjoint sensitivity method is a memory efficient alternative which is used in our work. The method description and its derivation are included in the supplemental material S2. {In this work, the neural network $\hat{f}$ is implemented using the PyTorch nn module, and the Adam optimizer is used to update the neural network parameters. Lastly, the adjoint sensitivity method is implemented as a custom autograd function for gradient propagation \cite{NeuralODE_github}}.

The ability to incorporate knowledge should require less training data for the learning framework to capture the true dynamics of the system of interest. And most importantly, the resulting hybrid model will be able to extrapolate beyond the sampled data, as we will demonstrate in our results.

\section{Results}

Using our universal learning framework, we consider different dynamical systems of varying complexities. For each system, training data is simulated with a suitable integration scheme. The simulation
and training parameters are summarized in Supplemental Material S3. Note that the training solver and its parameters must be chosen in a way such that it can realize the system’s trajectories. {Generally, solver selection should be based on criteria including system dynamics, solution stability, computation efficiency, and solver robustness \cite{solverchoice}.} For instance, Euler’s method with a large step size may not suffice for stiff systems such as the Van der Pol oscillator as it will result in numerical instability. The proposed framework is evaluated with respect to its ability to reproduce the dynamics of the actual system and to extrapolate or predict future observations. We also validate the framework for learning the dynamics of systems without prior knowledge and with limited prior knowledge.

To help better visualize the evolution of system states, we plot the trajectories of the Hopf normal form and the chaotic Lorenz system in gradient color, where their trajectories start in blue (RGBA: $[0.0, 0.0, 0.5, 1.0]$) and end in yellow (RGBA: $[0.9, 1.0, 0.0, 1.0]$).

\begin{figure}
    \centering
    \includegraphics[width=0.48\textwidth]{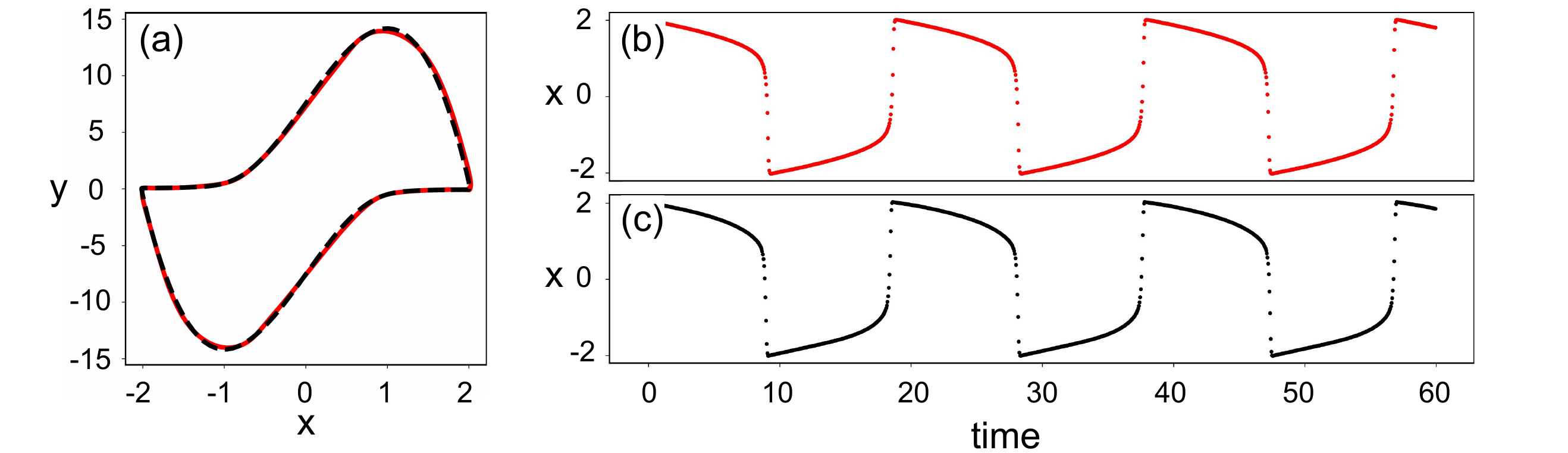}
    \caption{Prediction of the stiff Van der Pol oscillator with $\mu=10$. (a) Simulated trajectory (red) versus the model predicted trajectory (dotted black). (b) True trajectory, and (c) predicted trajectory of $x$ as a function of time for the Van der Pol oscillator.}
    \label{fig: vanderpol_result}
\end{figure}

\subsection{Learning Without Knowledge}
{In this subsection, we demonstrate learning without assuming any knowledge about the system, \textit{i.e.} only a neural network is used to model the systems and $\tilde{f} = \mathbf{0}$.}

\textbf{Van der Pol Oscillator.} We first demonstrate our framework in learning the dynamics of a stiff system. Consider the Van der Pol oscillator, which is a non-conservative oscillator with a limit cycle behavior~\cite{Enns1997} and has important applications in the modeling of periodic biological systems. The oscillator's second-order differential form is~\cite{vanderpol}
\begin{equation}
    \frac{d^2x}{dt^2} = \mu(1-x^2)\frac{dx}{dt}-x,
\end{equation}
but it is often written in its first-order form as the system of equations 
\begin{equation}
    \begin{aligned}
        \dot{x} &= y,\\
        \dot{y} &= -x +\mu (y-x^2y).
    \end{aligned}
\end{equation}

This system is known for its increasing \textit{stiffness} as $\mu$ is increased. Here, we consider the stiff Van der Pol oscillator with $\mu=10$. The training data is a trajectory of 1000 points simulated with the initial condition $[x_0, y_0]^\top = [2, 0]^\top$. The trained model correctly captures the stiffness and limit cycle behavior of the Van der Pol oscillator as shown in Fig. \ref{fig: vanderpol_result}.

\begin{figure}
    \centering
    \includegraphics[width=0.5\textwidth]{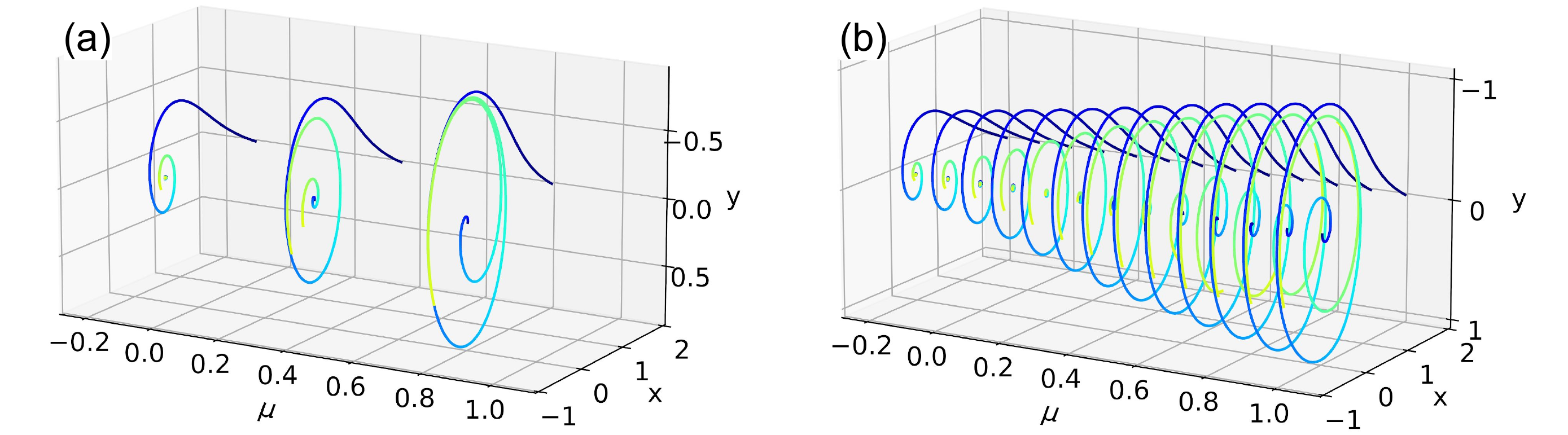}
    \caption{Learning the Hopf normal form. (a) Training data. (b) The trained model interpolates between $\mu=-0.2$ to $\mu=1.0$.}
    \label{fig:hopf}
\end{figure}

\textbf{Hopf Normal Form.} A bifurcation occurs when a dynamical system undergoes a sudden change in behavior when its parameters are varied. We consider a canonical model for a Hopf bifurcation, the Hopf normal form, which is given as
\begin{equation}
    \begin{aligned}
        \dot{x} &= \mu x + y - x(x^2+y^2),\\
        \dot{y} &= -x +\mu y - y (x^2+y^2),
    \end{aligned}
    \label{eqn: Hopf}
\end{equation}
where $\mu$ is the system parameter which, when varied, gives rise to a Hopf bifurcation. We incorporate the parameter $\mu$ into our model as a third dimension to the input vector, and set its time derivative to $0$, i.e., $\dot{\mu} = 0$.
Training data, as shown in \crefformat{figure}{Fig.~#2#1{(a)}#3}\cref{fig:hopf}, is simulated with three parameter values $\mu = \{-0.1, 0.35, 0.8\}$. The trained model correctly captures the bifurcation as shown in \crefformat{figure}{Fig.~#2#1{(b)}#3}\cref{fig:hopf}. The trained model can accurately interpolate between $\mu = -0.1$ to $\mu = 0.8$, and extrapolate from $\mu = -0.2$ to $\mu = 1.0$, which is beyond the range of $\mu$ in the training data. This demonstrates the generalization power of the trained model over system parameters.

\begin{figure}
    \centering
    \includegraphics[width=0.45\textwidth]{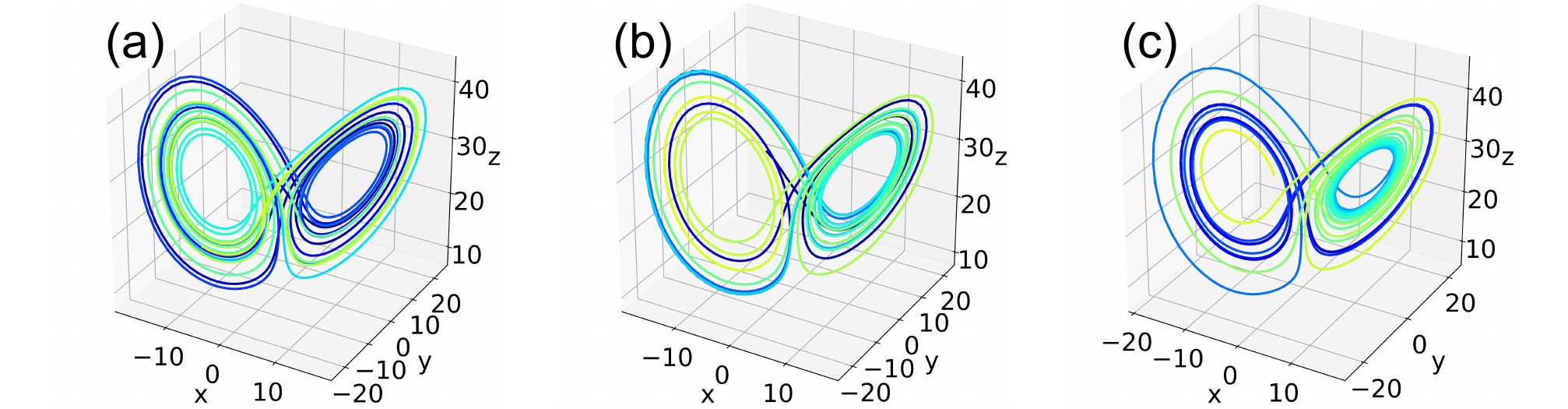}
    \caption{Learning the chaotic Lorenz system with a neural network without knowledge. (a) Training data ($t=0$ to $t=20$). (b) Model prediction ($t=0$ to $t=20$). (c) Model extrapolation ($t=20$ to $t=40$).}
    \label{fig:lorenz_no_knowledge}
\end{figure}

\textbf{Chaotic Lorenz System.} Next, we consider the chaotic Lorenz system~\cite{Lorenz}
\begin{equation}
    \begin{aligned}
        \dot{x} &= 10(y-x),\ \ 
        \dot{y} = x(28-z)-y,\\
        \dot{z} &= xy - (8/3)z.
    \end{aligned}
    \label{eqn: chaotic Lorenz}
\end{equation}
We simulate training data with the initial condition $[x(0), y(0),  z(0)]^\top = [-8, 7, 27]^\top$. As shown in Fig. \ref{fig:lorenz_no_knowledge}, the trained model is able to capture the bistable structure of the Lorenz attractor both within and beyond the duration of the training data.

To determine whether the resulting model is truly chaotic, we adopt the 0-1 test proposed by Gottwald and Melbourne~\cite{0-1}, which directly applies to time series data. The 0-1 test is binary, {\it i.e.}, under ideal conditions it outputs 1 if the system is chaotic and 0 otherwise. In practice, the outputs are approximately 1 and 0 for chaotic and non-chaotic systems. Details about the 0-1 test are included in the supplemental material S1.

Performing the 0-1 test on the predicted trajectory gives $K_c = 1.079$, showing that the identified system has chaotic dynamics.

\begin{figure}
        \includegraphics[width=0.45\textwidth]{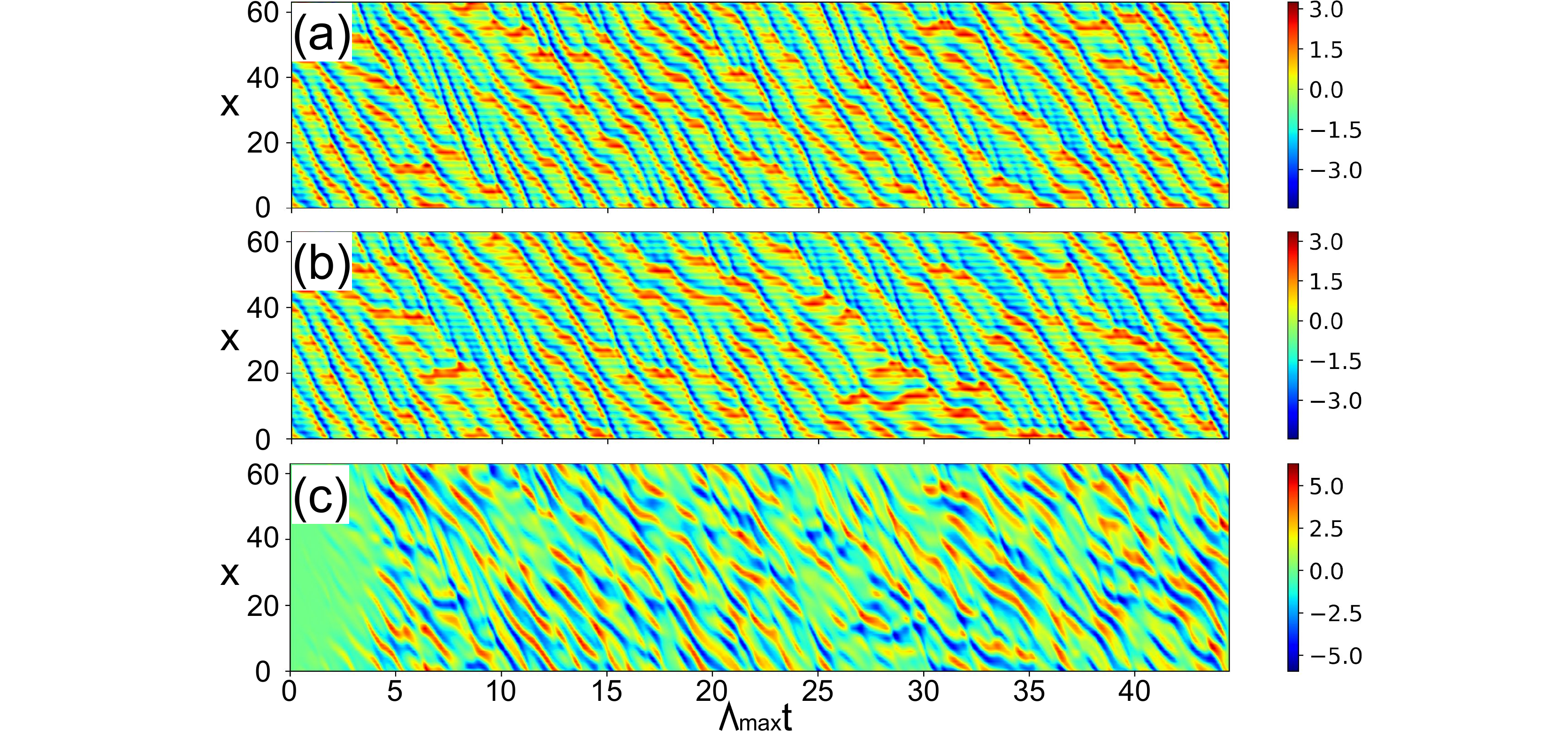}
        \caption{Trained KS model prediction on the testing data. (a) Testing data. (b) Prediction with the trained model using the first step of test data as the initial condition. (c) Difference between the test data and prediction.}
        \label{fig:KS extrapolation}
\end{figure}

\begin{figure}
        \includegraphics[width=0.43\textwidth]{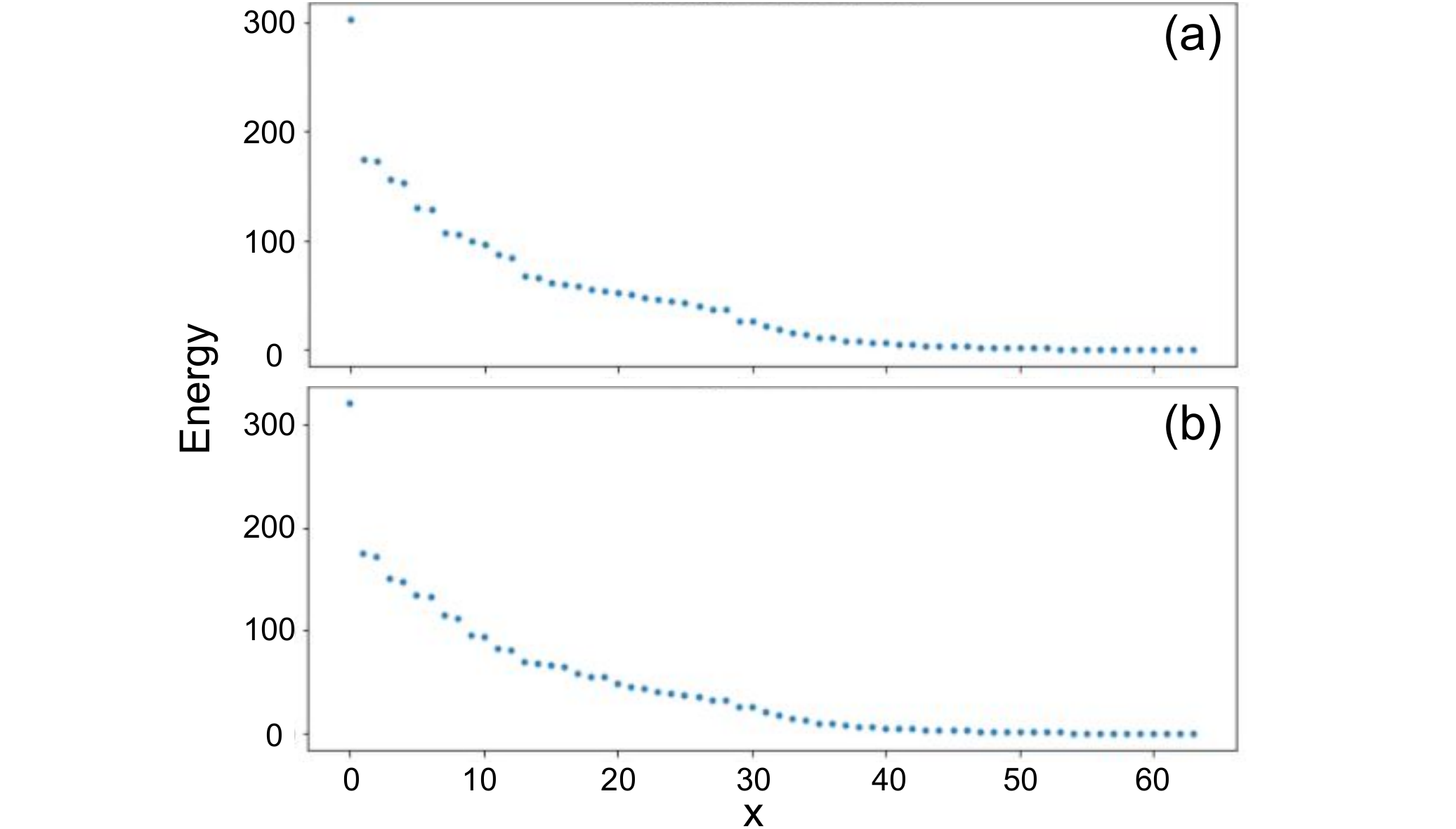}
        \caption{Comparing the POD decomposition for testing data and predictions. The scatter plots show the energy of the spatial modes from the POD decomposition of (a) the testing data, and (b) the predictions on the testing data.}
        \label{fig:KS POD}
\end{figure}

\textbf{Kuramoto-Sivashinsky Equation.} Moreover, our framework can learn the dynamics of spatiotemporally chaotic PDEs. Consider the one-dimensional \textit{Kuramoto-Sivashinsky equation} (KS) with periodic boundary conditions $y=y(x,t)$ and $y(x+L,t)=y(x,t)$~\cite{Kuramoto, Sivashinsky}
\begin{equation}
    \frac{\partial y}{\partial t}=-y\frac{\partial y}{\partial x} - \frac{\partial^2 y}{\partial^2 x} - \frac{\partial^4 y}{\partial^4 x}.
    \label{eqn: KS}
\end{equation}
We consider a 64-grid KS equation with $L = 60$. The most positive Lyapunov exponent is $\Lambda_{max} = 0.089$, and the Kaplan-Yorke dimension of the attractor is $D_{KY} = 13.56$~\cite{edson_bunder_mattner_roberts_2019}. We use a natural time scale, the Lyapunov time $\Lambda_{max} t$, for model evaluation. Since this system is spatially high-dimensional, we first use convolutional layers in our neural network for dimensionality reduction, and then a linear coupling matrix to restore the spatial dimension of the output. To generate training data, we discard the first 1000 steps of transient data and simulate a total of 8000 steps. Training data is then taken to be the first 5000 steps, which equates to 111.25 Lyapunov time. Testing data is taken from the 6000th to the 8000th steps of the simulated data. Fig. \ref{fig:KS extrapolation} shows that the trained neural network can capture the first 4 Lyapunov times with high accuracy before the trajectories diverge. Even after the trajectories diverge, the system behaves similarly to the actual system.

To qualitatively evaluate the performance of the trained model, we adopt the dimensionality reduction technique, proper orthogonal decomposition (POD)~\cite{holmes_lumley_berkooz_1996}, which maps the \textit{energy} of a given system. The term \textit{energy} is analogous to data variance in a principle component analysis~\cite{Jolliffe2016PrincipalCA}. Fig. \ref{fig:KS POD} shows the scatter plot of the \textit{energy} for each of the 64 dimensions in the testing data and the predictions made by the neural network.  It can be observed that the prediction has a similar energy signature as the testing data.

\begin{figure}
    \centering
    \includegraphics[width=0.48\textwidth]{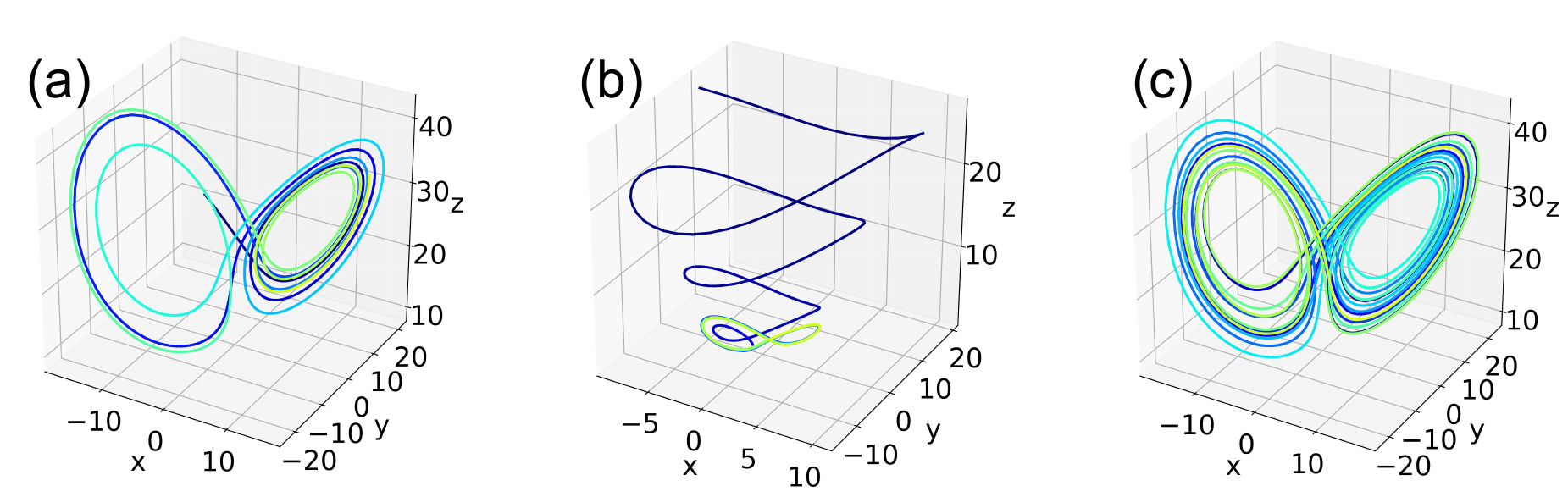}
    \caption{Learning the Lorenz system with knowledge. (a) Training data ($t=0$ to $t=8$). (b) Trajectory ($t=0$ to $t=20$) of the incorrect Lorenz system. (c) Extrapolation ($t=8$ to $t=28$) using trained model with the incorrect system as knowledge.}
    \label{fig:lorenz_with_knowledge}
\end{figure}

\subsection{Knowledge-Based Learning}
\begin{figure}
    \centering
    \includegraphics[width=0.5\textwidth]{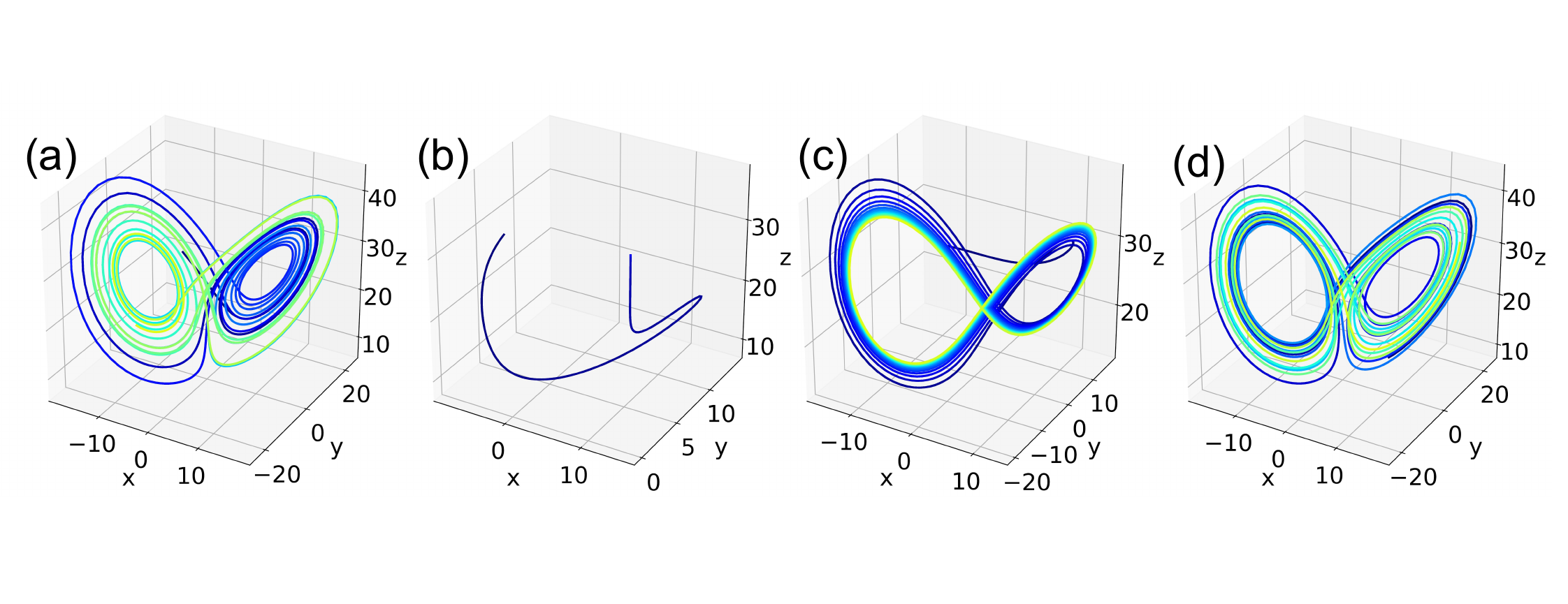}
    \caption{Systems identified by SINDy using a library of polynomial functions up to the fourth order after excluding different nonlinearities from the function library. Each plot shows the a trajectory for $t=0$ to $t=20$. (a) System identified if all terms are included in the library. (b) System identified if $x$, $xy$, and $xz$ are excluded from the library. (c) System identified if $xy$ and $xz$ are excluded from the library. (d) Extrapolation ($t=8$ to $t=28$) using trained model with the model in (c) as knowledge.}
        \label{fig: SINDy}
\end{figure}
{In this subsection, we consider learning with imperfect knowledge about a system. First, assume the true system is the Lorenz system from \eqref{eqn: chaotic Lorenz}. However, we only have the knowledge of a Lorenz system with incorrect coefficients.} The incorrect system keeps $\dot{x}$ and $\dot{z}$ the same but has $\dot{y} = x(-4.8-z)+7.2y$. {This incorrect system is therefore our knowledge $\tilde{f}$.} \crefformat{figure}{Fig.~#2#1{(b)}#3}\cref{fig:lorenz_with_knowledge} shows the trajectory of this incorrect system, which has periodic rather than chaotic behavior. We combine this incorrect system with a neural network according to the hybrid scheme shown in Fig. \ref{fig: hybrid_architecture}. The training data, shown in \crefformat{figure}{Fig.~#2#1{(a)}#3}\cref{fig:lorenz_with_knowledge}, is generated with the initial condition $[x(0), y(0), z(0)]^\top = [-8, 7, 27]^\top$. Note that this training data has only 2/5 the length used for training without knowledge. \crefformat{figure}{Fig.~#2#1{(c)}#3}\cref{fig:lorenz_with_knowledge} shows that the trained model restores the bistable structure and chaotic dynamics. Performing the 0-1 test on the knowledge gives $K_c = 0.036$, and the predicted trajectory gives $K_c = 1.11$, showing that the trained model has chaotic dynamics, even though the knowledge was not chaotic.

Next, we consider incorporating the knowledge generated by SINDy, a system identification method that uses sparse regression on a library of basis functions~\cite{SINDy}. When the correct nonlinear terms are missing from the library, SINDy may fail to capture the actual dynamics. Here we present an example of learning the chaotic Lorenz system using an imperfectly identified model from SINDy. We simulate observations for the chaotic Lorenz system using RK4 with a step size of $0.01$ from $t=0$ to $t=80$ and sample training data with a step size of $0.02$. We then perform SINDy on this data using a library of polynomials up to the fourth order. SINDy identifies the system shown in \crefformat{figure}{Fig.~#2#1{(a)}#3}\cref{fig: SINDy}, which correctly captures the bistable structure of the chaotic Lorenz system. However, if the nonlinearities $xz$ and $xy$ are excluded from the library of functions, SINDy identifies a system shown in \crefformat{figure}{Fig.~#2#1{(c)}#3}\cref{fig: SINDy}, which is not chaotic as the trajectory eventually forms a figure 8 shaped limit cycle. If $x$, $xz$, and $xy$ are excluded, the identified system, as shown in \crefformat{figure}{Fig.~#2#1{(b)}#3}\cref{fig: SINDy}, fails completely to capture the bistable structure.

The incorrectly identified system in \crefformat{figure}{Fig.~#2#1{(c)}#3}\cref{fig: SINDy} has the form

\begin{equation}
    \begin{aligned}
        \dot{x} &= -9.913x + 9.913y,\\
        \dot{y} &= -7.175x + 20.507y -0.613yz,\\
        \dot{z} &= -3.05z + 0.504x^2 + 0.479y^2,
    \end{aligned}
    \label{supp_eqn: SINDy_limit_cycle}
\end{equation}
{which we will use as the knowledge $\tilde{f}$.} Since the terms $xy$ and $xz$ are excluded from the regression library, both $\dot{y}$ and $\dot{z}$ have incorrect nonlinearities. By incorporating this incorrectly identified model according to the hybrid scheme, the trained model can correctly restore the bistable structure as shown in \crefformat{figure}{Fig.~#2#1{(d)}#3}\cref{fig: SINDy} {by only using training data from $t=0$ to $t=8$. Note again that the training data is only $2/5$ the length used for training without knowledge.} Performing the 0-1 test on the knowledge gives $K_c = 0.064$, and the predicted trajectory gives $K_c = 0.832$, showing that the trained model has chaotic dynamics.

We trained on the same amount of data ($t=0$ to $t=8$) without incorporating any knowledge. The neural network tends to overfit and "memorize" the data. When further extrapolating in time, the trained model tends to either a limit cycle or fixed point, and is unable to reproduce the chaotic dynamics.

\begin{figure}
    \centering
    \includegraphics[width=0.45\textwidth]{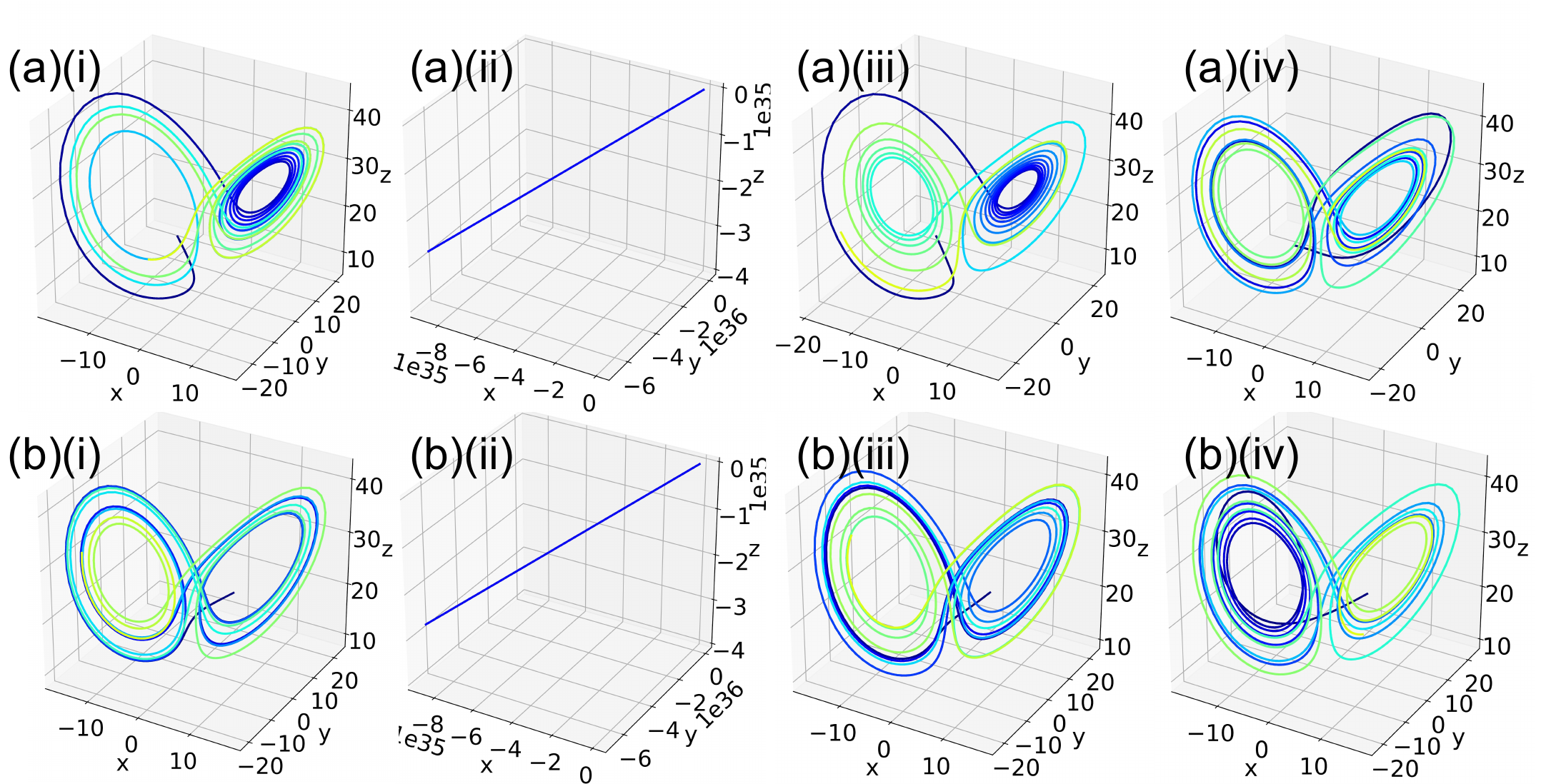}
    \caption{Model predictions from different initial conditions (t=0 to t=12). Trajectories are generated from the initial condition (a) $[-8,7,12]^\top$, and (b) $[10,-5,27]^\top$, using the (i) true Lorenz system, (ii) the neural network without knowledge, (iii), the neural network combined with incorrect Lorenz system, and (iv) the neural network combined with incorrectly identified model from SINDy.}
    \label{fig:extrapolation_comparison}
\end{figure}

Next, we show that knowledge incorporation gives the resulting model better extrapolation power beyond the sampled data. Here we compare the three trained models by using them to produce trajectories at initial conditions different from $[-8, 7, 27]^\top$, which is used for generating training data. Fig. \ref{fig:extrapolation_comparison} shows the predicted trajectories using the initial conditions $[-8, 7, 12]^\top$, and $[10, -5, 27]^\top$. While the chaotic behavior is completely lost in the neural network trained without any knowledge incorporation, both knowledge-based neural networks can still reproduce the bistable structure. Performing the 0-1 test on both knowledge-based models shows that they are chaotic.

\begin{figure}
    \centering
    \includegraphics[width=0.45\textwidth]{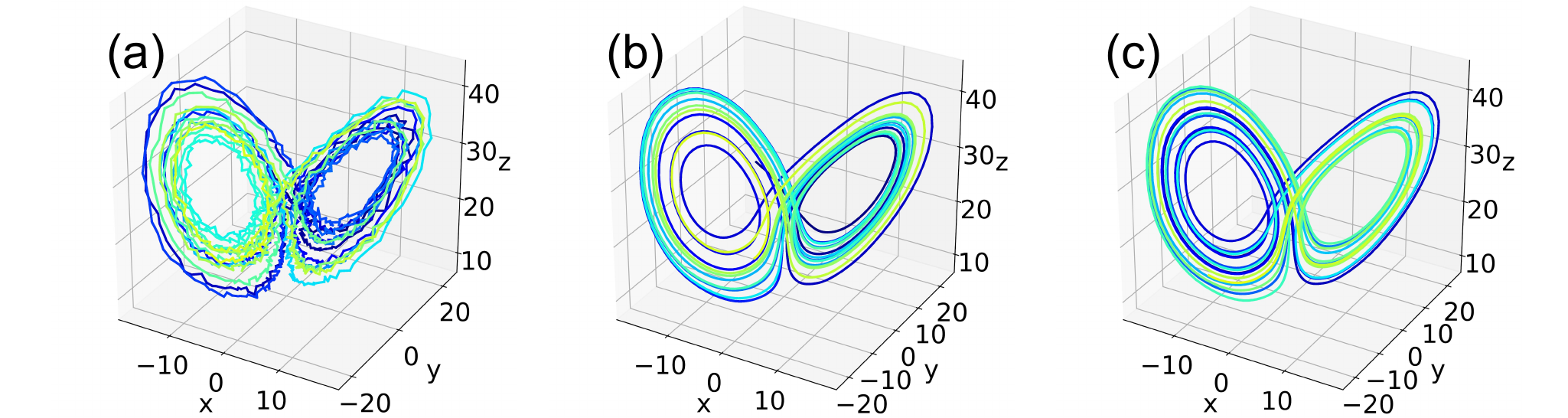}
    \caption{Learning the Lorenz system from noisy observations. (a) Training data ($t=0$ to $t=20$) with observational noise $\sim N(0, 0.1)$. (b) Model prediction ($t=0$ to $t=20$). (c) Model extrapolation ($t=20$ to $t=40$).}
    \label{fig: learning_with_noise}
\end{figure}

\begin{figure}
    \centering
    \includegraphics[width=0.45\textwidth]{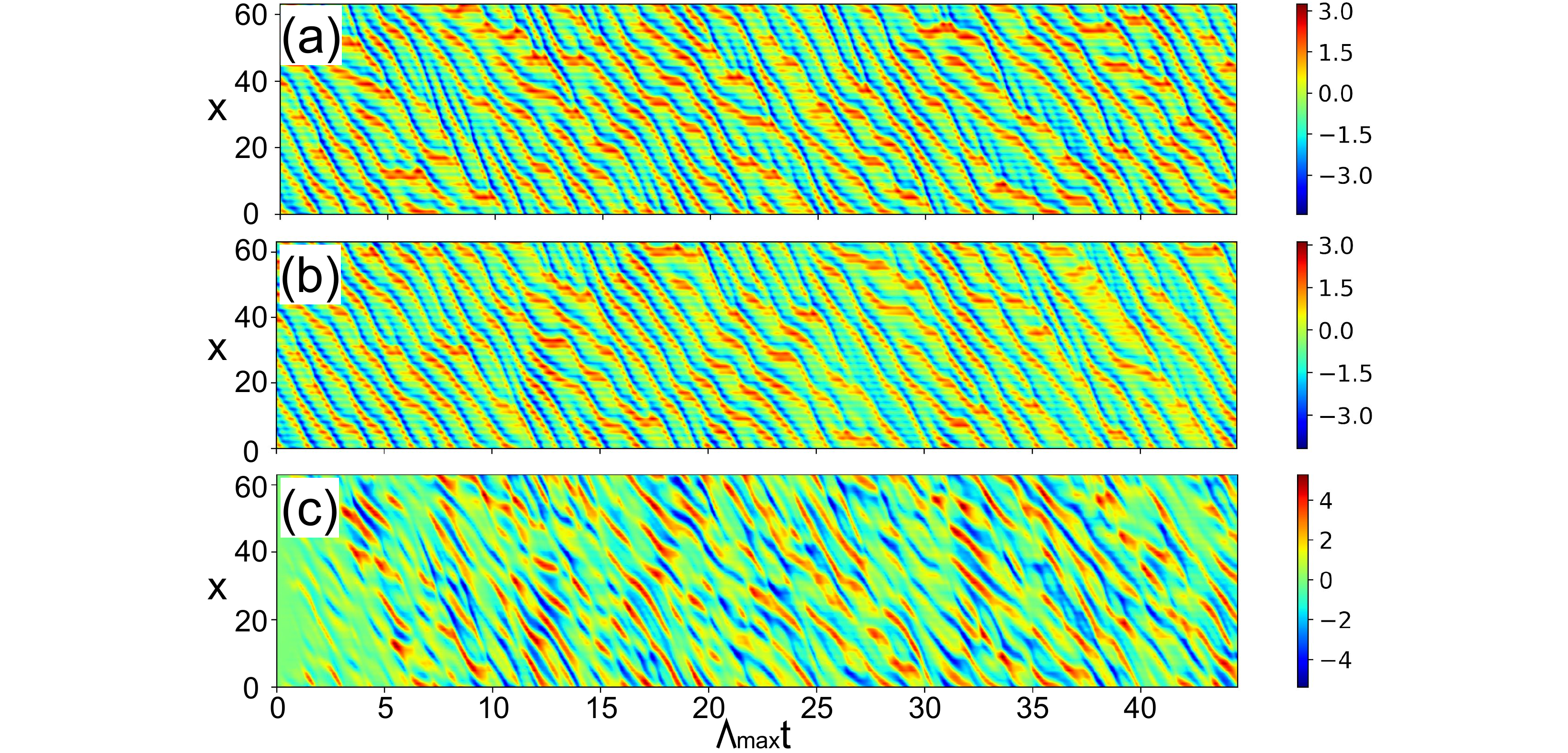}
    \caption{Learning from noisy observations of KS equation. (a) Testing data with $\sim N(0, 0.01)$ Gaussian noise. (b) Prediction with the trained model using the first step of test data as the initial condition. (c) Difference between the test data and prediction}
    \label{fig:KS noisy}
\end{figure}

\subsection{Learning from Noisy Observations}

Lastly, we demonstrate our framework's robustness to observational noise. Considering the Lorenz system, we use the same training data as training without knowledge but now include Gaussian noise with zero mean and 0.1 variance $(\sim N(0,0.1))$. This noisy trajectory is shown in \crefformat{figure}{Fig.~#2#1{(a)}#3}\cref{fig: learning_with_noise}. The trained neural network correctly captures the chaotic dynamics as shown in \crefformat{figure}{Fig.~#2#1{(b)}#3}\cref{fig: learning_with_noise} and \crefformat{figure}{~#2#1{(c)}#3}\cref{fig: learning_with_noise}. Performing the 0-1 test on the predicted trajectory gives $K_c = 1.10$.

We also consider noisy observations from the KS equation. Using the same training data as learning without knowledge, we include Gaussian noise with zero mean and 0.01 variance $(\sim N(0,0.01))$. Fig. \ref{fig:KS noisy} shows the learning result. Though the trained model can only accurately predict about 1 Lyapunov time before the trajectories diverge, this demonstrates the effect of noise on the chaotic dynamics, \textit{i.e.} a slight perturbation in the initial condition as a result of noise makes the trajectories diverge exponentially fast. However, it can be seen that the predicted trajectories behave similarly to the true system.

We note that the same neural network architecture converges much faster when trained on noisy observations. This is consistent with recent work that adds noise to observations of chaotic systems to stabilize the training process~\cite{OttRescomp}.

\section{Discussion}
We have demonstrated the universality of our framework by learning a wide variety of systems including stiff, bifurcating, and chaotic systems. We have also shown that with spatial convolution, our framework can easily scale and model high-dimensional systems. Note that since our framework does not require fixed time intervals for the training data, it can also learn from irregularly sampled trajectories (See Supplemental Material S6). {Most importantly, K-NODE is able to recover the dynamics of imperfect models by using them as knowledge. This demonstrates neural networks' capability to make up for the incorrect nonlinearities in the governing equations of systems.}

In this framework, the neural network is trained with respect to a numerical solver which can be different from the simulation solver used to generate the training data as demonstrated in learning the KS model. This means that we have the freedom to choose the training solver and its parameters depending on our needs. For example, we can use lower order solvers to speed up training at the expense of model accuracy, and we can perform finer temporal interpolations by using a smaller step size for the training solver at the expense of training speed. This is an important feature that ensures excellent accuracy without sacrificing computational speed. Moreover, it ensures that our framework can be applied not only in the case when one has high-fidelity simulation data, but also when one is working with real data, that may be noisy or have missing/sparse data.

For future work, we seek to explore more hybrid-learning approaches by incorporating different types of knowledge.   Moreover, besides using spatial convolutions for dimension reduction, we will investigate parallel learning using our framework to achieve even better scalability for very high-dimensional systems. Furthermore, in the hybrid learning scheme, we have full knowledge about the trained linear coupling matrix $\mathbf{M}_{out}$, which dictates how knowledge is combined with neural networks. This matrix $\mathbf{M}_{out}$ could potentially inform us of the correctness of assumptions made, as well as help us understand the role of neural networks in the learnt dynamics. In the future we hope to leverage this to extract additional physical insight into a wide range of dynamical systems. Given the amount of model and real data available to us, and considering the complexity of these systems, our universal learning framework can be used to understand meaningful correlations and establish new relationships between processes, thereby enhancing our knowledge.

The code that supports the findings of this study are available at \href{https://github.com/TomJZ/K-NODE}{github.com/TomJZ/K-NODE}.

\section*{Supplementary Material}
See \href{https://aip-prod-cdn.literatumonline.com/journals/content/cha/2021/cha.2021.31.issue-11/5.0065617/20211109/suppl/supplementary_material.pdf?b92b4ad1b4f274c70877518713abb28bc4cf650685dfb071af6195f4048a41ff509582803913bd4716126da586868433e0c8c264f8c3a0580aafd84466f067a4e6139c04e3f385d0ef387d5c5cc804c5fa5296221122578c28ae1a04a9843b7a70558e82d2b7b0b1a5c6e4805ef7fa01cf1397f0a2af40923918758dadf05b557a14ae3c6a798ebeb65fedaa597efaa8cc03b7}{\textcolor{blue}{supplementary material}} for simulation and training details, tips for choosing the lookahead, additional experimental results on learning from noisy or irregularly sampled data, derivations of the adjoint sensitivity method, and details of the 0-1 test for chaos.

\section*{Acknowledgement} This work was funded by the Office of Naval Research (ONR) Award No. 14-19-1-2253.

\bibliography{references}

\end{document}